\documentclass[conference]{IEEEtran}
\IEEEoverridecommandlockouts

\usepackage{cite}
\usepackage{amsmath,amssymb,amsfonts}
\usepackage{algorithmic}
\usepackage{graphicx}
\usepackage{textcomp}
\usepackage{xcolor}
\def\BibTeX{{\rm B\kern-.05em{\sc i\kern-.025em b}\kern-.08em
    T\kern-.1667em\lower.7ex\hbox{E}\kern-.125emX}}

\usepackage[english]{babel}
\usepackage{hyperref}
\usepackage{listings}
\usepackage{xcolor}
\lstdefinestyle{code}{
    backgroundcolor=\color{gray!20},
    breaklines=true,
    captionpos=b
}

\usepackage{epsfig}
\usepackage{subcaption}
\usepackage{calc}
\usepackage{amssymb}
\usepackage{amstext}
\usepackage{amsmath}
\usepackage{amsthm}
\usepackage{multicol}
\usepackage{pslatex}
\usepackage{apalike}
\usepackage{algorithm2e}
\usepackage[bottom]{footmisc}

\usepackage{makecell}

\usepackage{pifont}
\newcommand{\cmark}{\ding{51}}%
\newcommand{\xmark}{\ding{55}}%

\newcommand*\yescheck[1]{%
  \expandafter\newcommand\csname #1yescheck\endcsname{\textcolor{#1}{\cmark}}%
}
\yescheck{orange}
\yescheck{green}
\yescheck{red}

\newcommand*\nocheck[1]{%
  \expandafter\newcommand\csname #1nocheck\endcsname{\textcolor{#1}{\xmark}}%
}
\nocheck{blue}
\nocheck{green}
\nocheck{red}

\newcommand*\maybecheck[1]{%
  \expandafter\newcommand\csname #1maybecheck\endcsname{\textcolor{#1}{$\sim$}}%
}
\maybecheck{orange}
\maybecheck{green}
\maybecheck{red}

\usepackage{fontawesome}
\newcommand*\bansign[1]{%
  \expandafter\newcommand\csname #1bansign\endcsname{\textcolor{#1}{\faBan}}%
}
\bansign{red}
\bansign{green}

\newcommand*\bugicon[1]{%
  \expandafter\newcommand\csname #1bugicon\endcsname{\textcolor{#1}{\faBug}}%
}
\bugicon{red}
\bugicon{orange}

\newcommand*\shieldicon[1]{%
  \expandafter\newcommand\csname #1shieldicon\endcsname{\textcolor{#1}{\faShield}}%
}
\shieldicon{green}

\usepackage{comment}

\begin{document}

\title{The IoT Breaches your Household Again*\\
\thanks{In Proceedings of the 21st International Conference on Security and Cryptography (\href{https://secrypt.scitevents.org/Home.aspx?y=2024}{Secrypt 2024}), ISBN 978-989-758-709-2, ISSN 2184-7711, pages 475-482.
DOI: \href{https://doi.org/10.5220/0012767700003767}{10.5220/0012767700003767}
}
}

\author{\IEEEauthorblockN{1\textsuperscript{st} Davide Bonaventura}
\IEEEauthorblockA{\textit{Dipartimento di Matematica e Informatica} \\
\textit{Universit\`a di Catania}\\
Catania, Italy \\
d.bonaventura@studium.unict.it\\ 
\href{https://orcid.org/0009-0004-4463-7991}{orcid.org/0009-0004-4463-7991}}
\and
\IEEEauthorblockN{2\textsuperscript{nd} Sergio Esposito}
\IEEEauthorblockA{\textit{Information Security Group}\\
\textit{Royal Holloway} \\
\textit{University of London}\\
Egham, UK \\
sergio.esposito.2019@live.rhul.ac.uk\\
\href{https://orcid.org/0000-0001-9904-9821}{orcid.org/0000-0001-9904-9821}}
\and
\IEEEauthorblockN{3\textsuperscript{rd} Giampaolo Bella}
\IEEEauthorblockA{\textit{Dipartimento di Matematica e Informatica} \\
\textit{Universit\`a di Catania}\\
Catania, Italy \\
giamp@dmi.unict.it\\ 
\href{https://orcid.org/0000-0002-7615-8643}{orcid.org/0000-0002-7615-8643}}
}

\maketitle

\begin{abstract} 
Despite their apparent simplicity, devices like smart light bulbs and electrical plugs are often perceived as exempt from rigorous security measures. However, this paper challenges this misconception, uncovering how vulnerabilities in these seemingly innocuous devices can expose users to significant risks.
This paper extends the findings outlined in previous work, introducing a novel attack scenario. This new attack allows malicious actors to obtain sensitive credentials, including the victim's Tapo account email and password, as well as the SSID and password of her local network. Furthermore, we demonstrate how these findings can be replicated, either partially or fully, across other smart devices within the same IoT ecosystem, specifically those manufactured by Tp-Link.
Our investigation focused on the Tp-Link Tapo range, encompassing smart bulbs (Tapo L530E, Tapo L510E V2, and Tapo L630), a smart plug (Tapo P100), and a smart camera (Tapo C200). Utilizing similar communication protocols, or slight variants thereof, we found that the Tapo L530E, Tapo L510E V2, and Tapo L630 are susceptible to complete exploitation of all attack scenarios, including the newly identified one. Conversely, the Tapo P100 and Tapo C200 exhibit vulnerabilities to only a subset of attack scenarios.
In conclusion, by highlighting these vulnerabilities and their potential impact, we aim to raise awareness and encourage proactive steps towards mitigating security risks in smart device deployment.
\end{abstract}


\section{\uppercase{Introduction}}
\label{sec:introduction}
The digital revolution in Internet of Things (IoT) devices has led to ``smart'' devices becoming more and more an integral part of our daily lives. 
From smart home appliances to industrial sensors, IoT has unlocked a world of convenience, efficiency, and innovation. 
The number of IoT devices worldwide is forecast to almost double from 15.1 billion in 2020 to more than 29 billion IoT devices in 2030~\cite{statistics}.
The interconnectedness always brings forth significant security challenges that cannot be ignored.
Due to their often neglected security, IoT devices are typically preferred devices by attackers. 
On average, 54\% of organizations experience attempted cyberattacks targeting IoT devices every week. This indicates a 41\% increase in the average number of weekly attacks per organization targeting IoT devices compared to 2022~\cite{checkpoint-blog}.

More and more inexpensive IoT devices are designed without a security-first mindset~\cite{sternum-blog}~\cite{andrew-laughlin}, and their long lifecycles can expose them to evolving threats for years. 
The consequences of inadequate IoT security can be far-reaching. A compromised IoT device not only poses a risk to the privacy and safety of users but can also serve as a gateway to launch larger-scale attacks on critical infrastructures.

We observe that usually different devices produced by the same manufacturer, belonging to the same product line, e.g. Tapo, share parts of the firmware and application protocols used for communication.
Following these observations, this paper rests on the following research questions: 
\textit{(i) Do IoT devices from the same vendor share similar vulnerabilities? (ii) What consequences does this have on the end user's security, privacy and safety?}

\subsection{Contributions}
To answer the research questions we chose Tp-Link's IoT ecosystem as the target. 
Our experiments are focused on the following Tp-Link Tapo IoT devices.
\begin{enumerate}
    \item Tp-Link Tapo Smart Wi-Fi Light device, Multicolor (L530E)~\cite{tp-link-tapo-l530e}, 
    targeted by previous work, leading to the discovery of several vulnerabilities~\cite{secrypt23:smart-bulb-can-be-hacked}.
    \item Tp-Link Tapo Smart Wi-Fi Light Bulb, Dimmable (L510E V2)~\cite{tp-link-tapo-l510e}.
    \item Tp-Link Tapo Smart Wi-Fi Spotlight, Multicolor (L630)~\cite{tp-link-tapo-l630}.
    \item Tp-Link Tapo Mini Smart Wi-Fi Socket (P100)~\cite{tp-link-tapo-p100}.
    \item Tp-Link Tapo Pan/Tilt Home Security Wi-Fi Camera (C200)~\cite{tp-link-tapo-c200}.
\end{enumerate}

We found that the tested Tapo devices, part of Tp-Link's IoT ecosystem, use the protocols outlined in previous work~\cite{secrypt23:smart-bulb-can-be-hacked}, or its variants.
Consequently, we had the intuition that all attack scenarios described in previous work, or at least some of them, could most likely be exploited across all devices in the Tp-Link IoT ecosystem.

Hence, our findings regarding the tested Tp-Link devices can be summarised as follows:
\begin{itemize}
    \item The L510E V2 and the L630 use the same protocols as the L530E, thereby making all attack scenarios exploitable.
    \item Communications between the Tapo app and the C200 are secured via TLS encryption, limiting exploitation of the vulnerabilities.
    \item The configuration process of the P100 occurs over Bluetooth rather than Wi-Fi, restricting exploitability to attack scenarios that don't target association and configuration processes.
\end{itemize}

Additionally, we introduce a new attack scenario leveraging the first two vulnerabilities outlined in previous work~\cite{secrypt23:smart-bulb-can-be-hacked}. In this scenario, the attacker authenticates as the Tapo device to the Tapo app. As a result, the attacker can obtain the victim's Wi-Fi SSID and password, as well as her Tapo email and password.

\subsection{Ethics and Responsible Disclosure}
All experiments only involve resources owned by the authors of this work, including devices, Wi-Fi networks, accounts, emails, and passwords. No user or third-party data was accessed during the experiments. 

Tp-Link acknowledged the issues we responsibly reported through their Product Security Advisory (PSA)~\cite{tp-link-psa}. We actively collaborated with them, by testing the fixes and confirming the attack scenarios are no longer exploitable or do not give the attacker any advantage. 
Tp-Link confirmed that they already released the necessary fixes to address the vulnerabilities and that the changes do not affect the normal use and stability of the products.

\subsection{Paper Summary}
This document proceeds with a brief overview of relevant literature in the subsequent section (\S\ref{sec:rw}), followed by a concise summary of prior research (\S\ref{sec:resume}).
Subsequently, the new attack scenario is explained in detail (\S\ref{sec:attack_scenario_6}).
Then, for all devices covered by our study, a detailed description of the applicability or non-applicability of each attack scenario is provided(\S\ref{sec:impact}).
Ultimately, pertinent conclusions are derived (\S\ref{sec:conclusion}).

\section{\uppercase{Related Work}}\label{sec:rw}
This section delves into the related work within the field of IoT security. 

Nebbione et al.~\cite{fi12030055} delved into popular IoT protocols for data sharing and service discovery. They underscored the security risks posed by protocol limitations, device constraints, and vulnerabilities. Their conclusion emphasizes the need for enhancing service discovery protocols, implementing end-to-end security, and raising user awareness about IoT security risks.

In the work by Yaacoub et al.~\cite{YAACOUB2023280}, the authors underscore the importance of implementing proactive security measures in IoT systems, and highlight the limitations of traditional security methods. 
Their solution involves periodic ethical hacking simulations and penetration tests across various IoT components. In conclusion, the paper advocates for continuous training for all employees to make IoT systems more secure.

Unlike similar studies often focused on individual devices, Heiding et al.~\cite{HEIDING2023103067} conducted systematic penetration tests on 22 smart devices across different categories commonly found in connected homes.
As a result, a total of 17 vulnerabilities were uncovered and published as new CVEs. These vulnerabilities could grant attackers physical access to homes, posing significant risks to residents.

In the work by Akhilesh et al.~\cite{autom_pentest_fram}, the authors focus on enhancing the security of smart home-based IoT devices through automated penetration testing. Manual testing of IoT devices is labour-intensive and requires in-depth knowledge. To streamline this process, authors developed an automated penetration testing framework. Five smart home IoT devices were selected for testing, and common vulnerabilities were identified. 
The Tp-Link devices were found to be the most vulnerable, while the Google Home Mini was the most secure. The study concludes that the framework can be used by non-experts, contributing to improved IoT security and safer smart homes.

Researchers are also exploring various approaches to enhance the security levels of the IoT. For example, Hassija et al.~\cite{8742551} show how four different technologies, i.e., blockchain, fog computing, edge computing, and machine learning, can be used to increase the level of security in IoT, solving some of the main security issues present in the four layers in which an IoT application can be divided, which are sensing layer, network layer, middleware layer, and application layer. 
Finally, \textit{Salah} and \textit{Khan}~\cite{review_blockchain} present and survey major security issues for the IoT environment and show how blockchain can solve many of them.

\subsection{Previous Attacks on Tapo Bulbs}\label{sec:resume}
Previous work on Tapo L530E smart bulbs~\cite{secrypt23:smart-bulb-can-be-hacked} delineates the communication process between Tapo devices and the Tapo app, comprising three primary macro-steps: 
(1) \textit{Device Discovery} - allows the Tapo app to locate the Tapo device within the local network, and to get the Tapo device’s configuration;
(2) \textit{Tapo Symmetric Key Exchange Protocol (TSKEP)} - allows the Tapo app and the Tapo device to exchange a symmetric session key;
(3) \textit{Tapo device usage} - allows the user to use the Tapo device via the Tapo app, by sending get and set messages.

Within these macro-steps, authors identify and explain four vulnerabilities:
\begin{itemize}
    \item Vulnerability 1. \textit{Lack of authentication of the Tapo device with the Tapo app} allows an adjacent attacker to impersonate the Tapo device with the Tapo app during the TSKEP step.
    \item Vulnerability 2. \textit{Hard-coded, short shared secret} allows an adjacent attacker to obtain the secret for authentication during the \textit{Device Discovery} phase.
    \item  Vulnerability 3. \textit{Lack of randomness during symmetric encryption} allows an adjacent attacker to make the AES128-CBC scheme deterministic.
    \item  Vulnerability 4. \textit{Insufficient message freshness} allows an adjacent attacker to replay messages both to the Tapo device and the Tapo app.
\end{itemize}
These vulnerabilites were exploited by the authors in five attack scenarios, which we hereby summarise:

\begin{itemize}
    \item Attack Scenario 1, \textit{Fake Bulb Discovery Messages Generation,} that allows to discover Tapo devices within the network and serve false configurations to the Tapo app.
    \item Attack Scenario 2, \textit{Password Exfiltration from Tapo User Account,} that allows to get the password in cleartext of the user's Tapo account, and its associated email account in hash form.
    \item Attack Scenario 3, \textit{MITM Attack with a Configured Tapo L530E,} that allows to perform a Man-in-the-Middle attack and violate the confidentiality and integrity of all messages exchanged between the Tapo app and the Tapo device. This results in the exfiltration of the Tapo account password in cleartext, and the associated email account in hash form.
    \item Attack Scenario 4, \textit{Replay Attack with the Smart Bulb as Victim,} that allows to replay previously intercepted messages. If the adversary can observe the smart bulb's behaviour when the message arrives, they can infer the message's meaning and reuse it at will.
    \item Attack Scenario 5, \textit{MITM Attack with an Unconfigured Tapo L530E,} that allows to perform a Man-in-the-Middle attack and intercept traffic between the Tapo app and the Tapo device during configuration. As Tapo username and password, together with the Wi-Fi SSID and Wi-Fi password are sent in Base64 encoding during configuration, the adversary is able to exfiltrate all information.
\end{itemize}

Finally, the authors conduct experiments across three different network setups, denoted as \textit{Setup A}, \textit{Setup B}, and \textit{Setup C}. In Setup A, both the victim (i.e., a phone running the Tapo app) and the adversary are connected to the same network, while the Tapo device is on a separate, remote network; in Setup B, the adversary, the victim and the Tapo device are all connected to the same local network, and the Tapo device is already configured; in Setup C, the adversary keeps deauthenticating~\cite{bellardo2003802} the Tapo device, resetting it to the unconfigured state, until the user connects it to the adversary's Wi-Fi honeypot, thinking it's their home network.

\section{\uppercase{Breaching The Household Again}}\label{sec:attack_scenario_6}

In this section, we present a novel attack scenario, which we call \textit{``Attack Scenario 6 - Passwords exfiltration with an unconfigured Tapo device''}, following the enumeration within previous work on Tapo devices~\cite{secrypt23:smart-bulb-can-be-hacked}. In this new attack scenario, the adversary is able to exfiltrate passwords using an unconfigured Tapo device.

The devices used during the attack are:\begin{itemize}
    \item A Wi-Fi switch to provide local connectivity.
    \item A smart bulb Tapo series L530 with Hardware Version 1.0.0 and Firmware Version 1.1.9.
    \item A Samsung smartphone running Android 11 and the Tapo app Version 2.8.14.
    \item An Ubuntu 22.04 machine with 5.15.0-47 kernel.
\end{itemize}

\subsection{Setup D}\label{sec:setup}
The network configuration we use during the attack, which we call \textit{Setup D}, for consistency with previous work~\cite{secrypt23:smart-bulb-can-be-hacked}, is as follows. 
\begin{itemize}
    \item The victim wants to associate an unconfigured Tapo device with her Tapo account.
    \item The Tapo app (hence, the victim) believes to be connected to the network $X$ created by the Tapo device, but is actually connected to a network $Y$ controlled by the attacker's Ubuntu device.
\end{itemize}

\begin{figure}[htbp]
  \centering
   {\epsfig{file = 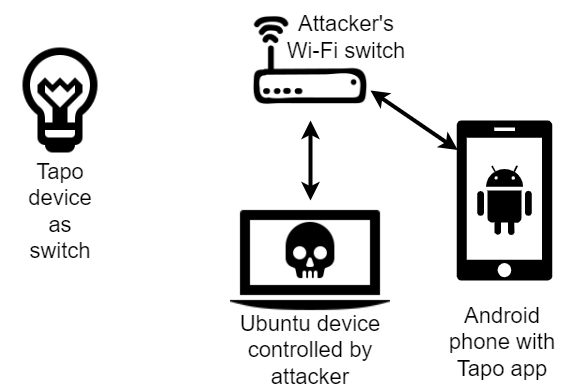, scale=0.35}}
  \caption{Setup D, network with a non-configured device.}
  \label{fig:setup}
\end{figure}

This setup requires that the Tapo device has been reset or has not been configured yet. 
The attacker must only be connected to the network they control, and not to the access point started by the Tapo device. The victim's Tapo app must be connected to the network controlled by the attacker.
In this setup, the victim opens the Tapo application and starts the device association process.
The network configuration for this setup is shown in Figure~\ref{fig:setup}.

As shown in previous work~\cite{secrypt23:smart-bulb-can-be-hacked}, the attacker can use the \textit{Wi-Fi deauthentication attack}~\cite{bellardo2003802} to easily get the \textit{Setup D} as well.
Initially, the adversary can use the deauthentication attack to disconnect the Tapo device from the network to which it is connected, forcing the victim to reset it.
Subsequently, after the Tapo device enters setup mode, the attacker can perform the same attack to deauthenticate the Tapo app from the network started by the Tapo device, trying to get the victim to connect to the network they control.

\subsection{Attack Scenario 6}
\begin{figure}[htbp]
  \centering
   {\epsfig{file = 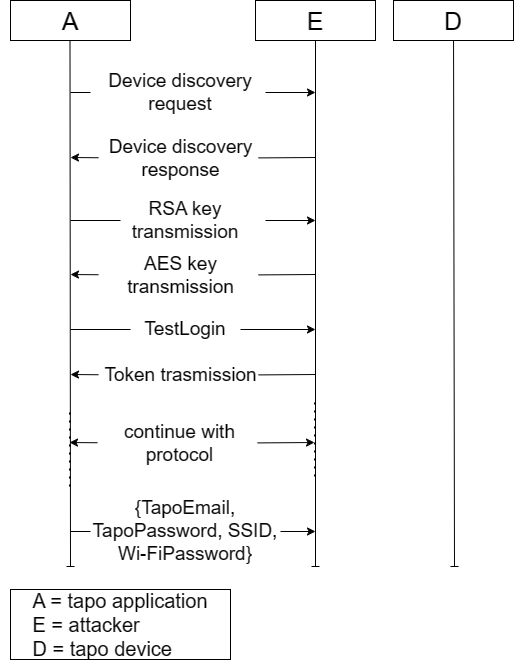, scale=0.40}}
  \caption{Sequence chart for the attack scenario 6.}
  \label{fig:attack_6}
\end{figure}

In this experiment, we exploited two of the four vulnerabilities, in order:
\begin{itemize}
    \item \textit{Vulnerability 2}, with the goal of creating fake \textit{device discovery response},
    \item \textit{Vulnerability 1}, with the goal of authenticating as the Tapo device to the Tapo app.
\end{itemize}
The context in which we conduct the experiment is the Setup D(\S\ref{sec:setup}) previously described.
The attack diagram is shown in Figure~\ref{fig:attack_6}.

The exploitation begins when the victim starts the association process within the Tapo app. In the beginning, the app starts broadcasting \textit{device discovery request}. 
Hence, the attacker exploits his ability to create fake \textit{device discovery response} to respond to various \textit{device discovery request} from the victim. The attacker sets the response's messages fields as shown in Listing \ref{lst:attacco 6}:
\begin{itemize}
    \item He sets the \textit{device\_id} and \textit{owner} fields with random hex values.
    \item He sets the \textit{device\_type} and \textit{device\_model} fields with the name and type of the device he wants to impersonate. 
    \item He sets the \textit{ip} and \textit{port} fields to point to an adversary-controlled server.
    \item He sets the \textit{factory\_default} field to \texttt{true}. This is important because it allows the application to understand that the response is coming from a device not yet associated with any accounts.
\end{itemize}
Note that, differently from the \textit{Attack Scenario 2} described in previous work~\cite{secrypt23:smart-bulb-can-be-hacked}, the attacker does not need the victim's \textit{owner id}.

\begin{small}
\begin{lstlisting}[style=code, caption={JSON attack scenario 6},label={lst:attacco 6}]
{
  "result": {
    "device_id": "RANDOM.HEX.VALUE",
    "owner": "RANDOM.HEX.VALUE",
    "device_type": "DEVICE.TYPE",
    "device_model": "DEVICE.MODEL",
    "IP": "ATTACKER.IP",
    "mac": "ATTACKER.PORT",
    "factory_default": true,
    "is_support_iot_cloud": false,
    "mgt_encrypt_schm": {
      "is_support_https": false,
      "encrypt_type": "AES",
      "http_port": 80
    }
  },
  "error_code": 0
}
\end{lstlisting}
\end{small}

After receiving the response, the Tapo app assumes that it comes from a device that needs to be associated. Therefore, it starts the TSKEP protocol with the attacking device. 
Because of \textit{vulnerability 1}, the TSKEP protocol does not give the Tapo app any evidence about the identity of the interlocutor. For this reason, the Tapo app assumes that the newly received key is shared with the device to be associated, while it is shared with the attacker instead. 
The attacker must then perform the association process with the Tapo app until the \textit{set\_qs\_info} request. At that point, they can get the password of the victim's Tapo account and the associated email address, as well as the SSID and the password of the victim's local network.

The attack can be summarised as follows:\begin{itemize}
        \item The attacker gets the \textit{Device Discovery} shared key and creates fake \textit{device discovery response}. Therefore, the authentication of the \textit{device discovery response} fails. 
        
        \item The Tapo app executes the TSKEP protocol with the attacker instead of the Tapo device. Therefore, authentication of the Tapo device with the Tapo app fails. This results in an \textit{integrity loss}.
        
        \item The Tapo app shares the key with the attacker, hence the distribution of the session key fails. This results in an \textit{confidentiality loss}.

        \item The attacker can violate the confidentiality of the messages and get the password and the hash of the email of the victim’s Tapo account as well as the SSID and the password of the victim's local network. This results in a \textit{confidentiality loss}.
\end{itemize}

\section{\uppercase{Impact on Target Devices}}\label{sec:impact}
Table~\ref{tab:firmware_versions} provides an overview of the different versions tested for Tapo devices and the Tapo app. We tested versions of device firmwares that were supposedly vulnerable to the discovered vulnerabilities, then we tested the fixed firmwares to check if the vulnerabilities were not exploitable anymore.

\begin{table}[htbp]
\normalsize
\begin{center}
\caption{Tested versions}
\begin{tabular}{ c c c }
\Xhline{2\arrayrulewidth}
 & \textbf{Vulnerable Version} & \textbf{Fixed Version} \\
\Xhline{2\arrayrulewidth}
L530E & 1.1.9 & 1.2.4 \\ \hline
L510E & 1.0.8 & 1.1.0 \\ \hline
P100 & 1.4.9 and  1.4.16 & 1.5.0 \\ \hline
C200 & 1.1.18 & - \\ \hline
L630 & 1.0.3 & 1.0.4 \\ \hline
Tapo App & 2.8.14 and 2.16.112 & 2.17.206 \\ \Xhline{2\arrayrulewidth}
\end{tabular}
\label{tab:firmware_versions}
\end{center}
\end{table}

\subsection{Firmware Without Fixes}
Our primary focus is to assess the impact of the identified vulnerabilities on each target device. A summary of the vulnerabilities exposed by each target device running a firmware without fixes is shown in Table~\ref{tab:firmwarevuln_without_fixes}. 
Throughout the section, we also analyze the applicability of the attack scenarios, and we summarise the reproducible ones on each target device in Table~\ref{tab:firmware_without_fixes}.

\begin{table}[htbp]
\normalsize
\begin{center}
\caption{Vulnerabilities exposed by target devices for firmware without fixes}
\begin{tabular}{ c c c c c }
\Xhline{2\arrayrulewidth}
& \textbf{Vuln. 1} & \textbf{Vuln. 2} & \textbf{Vuln. 3} & \textbf{Vuln. 4} \\
\Xhline{2\arrayrulewidth}
\textbf{L530E} & \redbugicon & \redbugicon & \redbugicon & \redbugicon \\ \hline
\textbf{L510E} & \redbugicon & \redbugicon & \redbugicon & \redbugicon \\ \hline
\textbf{L630} & \redbugicon & \redbugicon & \redbugicon & \redbugicon \\ \hline
\textbf{P100} & \redbugicon & \redbugicon & \redbugicon & \redbugicon \\ \hline
\textbf{C200} & \small{\greenshieldicon} & \redbugicon & \small{\greenshieldicon} & \small{\greenshieldicon} \\
\Xhline{2\arrayrulewidth}
\end{tabular}\\ 
\footnotesize{\redbugicon\ if the vulnerability is present, \small{\greenshieldicon}\ otherwise.}
\label{tab:firmwarevuln_without_fixes}
\end{center}
\end{table}

\begin{table}[htbp]
\normalsize
\begin{center}
\caption{Feasibility of the Attack Scenarios (AS) on the target devices for firmware without fixes}
\begin{tabular}{ c c c c c c c }
\Xhline{2\arrayrulewidth}
 & \textbf{AS1} & \textbf{AS2} & \textbf{AS3} & \textbf{AS4} & \textbf{AS5} & \textbf{AS6} \\ \Xhline{2\arrayrulewidth}
\textbf{L530E} & \redbugicon & \redbugicon & \redbugicon & \redbugicon & \redbugicon & \redbugicon \\ \hline
\textbf{L510E} & \redbugicon & \redbugicon & \redbugicon & \redbugicon & \redbugicon & \redbugicon \\ \hline
\textbf{L630} & \redbugicon & \redbugicon & \redbugicon & \redbugicon & \redbugicon & \redbugicon \\ \hline
\textbf{P100} & \redbugicon & \redbugicon & \redbugicon & \redbugicon & \greenbansign & \greenbansign \\ \hline
\textbf{C200} & \redbugicon & \small{\greenshieldicon} &  \small{\greenshieldicon} &  \small{\greenshieldicon} &  \small{\greenshieldicon} &  \small{\greenshieldicon} \\ 
\Xhline{2\arrayrulewidth}
\end{tabular} \\
\footnotesize{\redbugicon\ if the attack scenario is feasible on the target device, \small{\greenshieldicon}\ if the attack scenario is not feasible because the communication is encapsulated within a TLS channel, \greenbansign\ if the attack scenario is not feasible because the configuration is done on the Bluetooth channel.} \\ 
\label{tab:firmware_without_fixes}
\end{center}
\end{table}

\subsubsection{Tp-Link Tapo Smart Wi-Fi Light device, Multicolor (L530E)}
We test an L530E with \textit{Hardware v1.0.0} and \textit{Firmware v1.1.9}, using a Tapo app \textit{v2.8.14}. 
The smart bulb exposes all the vulnerabilities, and all \textit{attack scenarios} are reproducible~\cite{secrypt23:smart-bulb-can-be-hacked}, including our novel \textit{Attack Scenario 6} (\S\ref{sec:attack_scenario_6}).

\subsubsection{Tp-Link Tapo Smart Wi-Fi Light Bulb, Dimmable (L510E V2)}
We test an L510E V2 with \textit{Hardware v2.0} and \textit{Firmware v1.0.8}, using a Tapo app \textit{v2.16.112}.
The Tapo L510E, for the \textit{Device-App} communications, uses the same vulnerable protocols (\S\ref{sec:resume}) with the same security parameters used by the L530E, i.e., HTTPS protocol is not supported, CBC-AES128 bit encryption is used, and Wi-Fi is the communication channel during configuration. 
Therefore, this smart bulb exposes all the listed vulnerabilities, and all  \textit{attack scenarios} can be reproduced, including \textit{Attack Scenario 6} introduced in this paper. 

We hereby describe how we apply each attack scenario to the new device, highlighting any differences from previous work~\cite{secrypt23:smart-bulb-can-be-hacked} as necessary. For newly tested devices, we will refer to the L510E's behaviour as a baseline. In later sections, we will only detail Attack Scenarios (AS) that deviate from this baseline.

\begin{description}
   \item[AS1] works with the Tapo L510E firmware tested. 
   The key that this device uses for the Message Authentication Code is static and hardcoded, the same used by the Tapo L530E. Therefore, an attacker can create false \textit{device discovery} messages for both the bulb and the app.

   \item[AS2] works with the Tapo L510E firmware tested.
   The Tapo L510E communicates using the TSKEP protocol with the Tapo app. By creating fake \textit{device discovery response}, the attacker can impersonate the Tapo L510E, prompting the app to start TSKEP with them. This allows the attacker to get the Tapo password and the hash of the victim's Tapo email.
   
   \item[AS3] works with the Tapo L510E firmware tested.
   TSKEP lacks identity verification, enabling the attacker to perform a MITM attack on the Tapo L510E-Tapo app communication, compromising confidentiality.
   
   \item[AS4] works with the Tapo L510E firmware tested.
   The Tapo L510E accepts all messages without checking their timestamp. This allows attackers to replay sniffed messages with non-expired session keys, enabling arbitrary command execution.
   
   \item[AS5] works with the Tapo L510E firmware tested.
   During pairing, communications between the Tapo L510E and the Tapo app happen over Wi-Fi. Hence, the attacker can perform a MITM attack and hijack the association process.
   
   \item[AS6] works with the Tapo L510E firmware tested.
   During pairing, Tapo L510E and Tapo app communicate via Wi-Fi. TSKEP's identity verification vulnerability allows MITM attacks, compromising the email and password of the victim's Tapo account, as well as the SSID and password of her local network.
\end{description}

\subsubsection{Tp-Link Tapo Smart Wi-Fi Spotlight, Multicolor (L630)}
We test an L630 with \textit{Hardware v1.0} and \textit{Firmware v1.0.3}, using a Tapo app \textit{v2.16.112}.
We confirmed this device aligns with our baseline —-- mirroring the behavior of the Tapo L510E V2. Thus, it shares all listed vulnerabilities and allows reproduction of all \textit{attack scenarios}, including the new \textit{Attack Scenario 6} introduced in this paper.

\subsubsection{Tp-Link Tapo Mini Smart Wi-Fi Socket (P100)}
We test a P100 with \textit{Hardware v1.20.0} and \textit{Firmwares v1.4.9 and v1.4.16}, using Tapo app \textit{v2.16.112}.
This device employs vulnerable protocols (\S\ref{sec:resume}), lacks HTTPS support, and uses CBC-AES128 encryption, exposing all vulnerabilities. Unlike previous devices, P100 uses Bluetooth for configuration, limiting attack scenarios to those involving already associated devices. Hence, Attack Scenarios 1 to 4 are aligned with our baseline,
while Attack Scenarios 5 and 6 cannot be reproduced on Tapo P100 because the adversary is not able to perform the MITM attack during the bulb configuration process.

\subsubsection{Tp-Link Tapo Pan/Tilt Home Security Wi-Fi Camera (C200)}
We test a C200 with \textit{Hardware v1.0.0} and \textit{Firmware v1.1.18}, using a Tapo app \textit{v2.16.112}.

Unlike the other analysed devices, the Tapo C200 supports HTTPS, utilizing TLS for TSKEP between the Tapo app and device even during configuration. This limits exposure to only \textit{Vulnerability 2}. The use of TLS prevents message inference or traffic sniffing by requiring a valid certificate from the attacker. While TSKEP remains vulnerable to replay attacks, TLS encapsulation ensures security. Consequently, only \textit{Attack scenario 1} can be reproduced out of six attack scenarios.

One potential attack involves downgrading the communication channel from HTTPS to HTTP. The attacker may attempt this by replying to the \textit{device discovery requests} from the application with the same security parameters supported by the Tapo L530E, i.e., HTTPS not supported, as shown in Listing~\ref{lst:c200_sub_attack}.
However, we verified that this downgrade attack produces no results. This is because the Tapo application does not consider valid all \textit{device discovery response} received from C200 devices that do not support HTTPS.
\begin{small}
\begin{lstlisting}[style=code,caption={Attack sub-scenario 2 UDP discovery response},label={lst:c200_sub_attack}]
{
  "error_code": 0,
  "result": {
    "device_id": "1234...441",
    "device_name": "Tapo_Camera_E3FF",
    "device_type": "SMART.IPCAMERA",
    "device_model": "C200",
    "ip": "192.168.1.55",
    "mac": "AA-BB-CC-DD-EE-FF",
    "hardware_version": "1.0",
    "firmware_version": "1.1.18 Build 220518 Rel.61472n(4555)",
    "factory_default": false,
    "is_support_iot_cloud": false,
    "mgt_encrypt_schm": {
      "is_support_https": false,
      "encrypt_type": "AES",
      "http_port": "Evil.tcp_port"
    }
  }
}
\end{lstlisting}
\end{small}
    
\subsection{Firmware With Fixes}
For each device tested, we diligently communicated the discovered vulnerabilities to Tp-Link. The responsible disclosure process enabled Tp-Link to promptly identify and address the vulnerabilities. They developed new versions of the Tapo app and the Tapo devices' firmware, implementing security updates to resolve the issues. We then actively tested the beta versions of this firmware, confirming the mitigation of potential risks arising from the vulnerabilities, and providing feedback to the manufacturer. 

Although only three out of the four vulnerabilities, i.e., \textit{Vuln. 1}, \textit{Vuln. 3}, and \textit{Vuln. 4}, were addressed with fixes, their absence indirectly mitigates the risk associated with the remaining vulnerability, i.e., \textit{Vuln. 2}, making it acceptable. Therefore, even if the last vulnerability is still exposed, it would not pose a significant security risk to the end user.
A summary of the vulnerabilities exposed by each target device running a firmware with fixes is shown in Table~\ref{tab:firmwarevuln_with_fixes}.
\begin{table}[htbp]
\normalsize
\begin{center}
\caption{Vulnerabilities exposed by target devices for firmware with fixes}
\begin{tabular}{ c c c c c }
\Xhline{2\arrayrulewidth}
 & \textbf{Vuln. 1} & \textbf{Vuln. 2} & \textbf{Vuln. 3} & \textbf{Vuln. 4} \\
 \Xhline{2\arrayrulewidth}
\textbf{L530E} & \small{\greenshieldicon} & \redbugicon & \small{\greenshieldicon} & \small{\greenshieldicon} \\ \hline
\textbf{L510E} & \small{\greenshieldicon} & \redbugicon & \small{\greenshieldicon} & \small{\greenshieldicon} \\ \hline
\textbf{L630} & \small{\greenshieldicon} & \redbugicon & \small{\greenshieldicon} & \small{\greenshieldicon} \\ \hline
\textbf{P100} & \small{\greenshieldicon} & \redbugicon & \small{\greenshieldicon} & \small{\greenshieldicon} \\ \hline
\textbf{C200} & - & \redbugicon & - & - \\
\Xhline{2\arrayrulewidth}
\end{tabular} \\
\footnotesize{\redbugicon\ if the vulnerability is still present, \small{\greenshieldicon}\ otherwise,\\ - if it was not present in the unpatched firmware.}
\label{tab:firmwarevuln_with_fixes}
\end{center}
\end{table}

Regarding the attack scenarios, we tested all six of them using the beta version of the Tapo app, specifically \textit{Version 2.17.206}, and the device's firmware provided by the Tp-Link. Only one of the six attack scenarios can still be reproduced, i.e., \textit{Attack scenario 1, Fake Bulb Discovery Messages Generation}.
However, the inability for the adversary to reproduce the other scenarios renders Attack scenario 1 virtually negligible in terms of risk to the victim, thus offering no advantage to the potential attacker. This observation confirms that all attack scenarios are effectively nullified, as none yields any results.
A summary of the reproducible attack scenarios on each device running a firmware with fixes is shown in Table~\ref{tab:firmware_with_fixes}.
\begin{table}[htbp]
\normalsize
\begin{center}
\caption{Impact of the Attack Scenarios (AS) on the target devices for firmware with fixes}
\begin{tabular}{ c c c c c c c }
\Xhline{2\arrayrulewidth}
 & \textbf{AS1} & \textbf{AS2} & \textbf{AS3} & \textbf{AS4} & \textbf{AS5} & \textbf{AS6} \\ \Xhline{2\arrayrulewidth}
\textbf{L530E} & \orangebugicon & \small{\greenshieldicon} & \small{\greenshieldicon} & \small{\greenshieldicon} & \small{\greenshieldicon} & \small{\greenshieldicon} \\ \hline
\textbf{L510E} & \orangebugicon & \small{\greenshieldicon} & \small{\greenshieldicon} & \small{\greenshieldicon} & \small{\greenshieldicon} & \small{\greenshieldicon} \\ \hline
\textbf{L630} & \orangebugicon & \small{\greenshieldicon} & \small{\greenshieldicon} & \small{\greenshieldicon} & \small{\greenshieldicon} & \small{\greenshieldicon} \\ \hline
\textbf{P100} & \orangebugicon & \small{\greenshieldicon} & \small{\greenshieldicon} & \small{\greenshieldicon} & - & - \\ \hline
\textbf{C200} & \orangebugicon & - & - & - & - & - \\ 
\Xhline{2\arrayrulewidth}
\end{tabular} \\
\footnotesize{\orangebugicon\ if the AS is feasible without benefits for the attacker,\\ \small{\greenshieldicon}\ if the AS is not feasible anymore, \\ - if the AS was not feasible in the unpatched firmware.}
\label{tab:firmware_with_fixes}
\end{center}
\end{table}

\section{\uppercase{Conclusions}}\label{sec:conclusion}
In this paper, we attempted to exploit different Tapo devices using vulnerabilities that affected the Tapo L530E smart bulb, which were found in previous work. Results show that said vulnerabilities are present and exploitable in other devices belonging to the Tp-Link ecosystem and not exclusive to a specific Tapo device. More generally, to answer our first research question, this hints at the fact that the stack of technologies underlying IoT devices is shared between devices of the same family, and that advisories published for a single device may actually be helpful to both attackers and defenders in identifying the same vulnerabilities on other devices of the same ecosystem. This is most likely not unique to the Tapo environment, but verification of this claim is left to future work.

Additionally, we expanded previous work by introducing a new Attack Scenario, which we called \textit{``Attack Scenario 6''}, and a novel network configuration to exploit the vulnerabilities, which we called \textit{``Setup D''}. We then tested all attack scenarios on different Tapo devices, finding that they are mostly reproducible, with a few exceptions.
Hence, we answer our second research question by verifying that exploitable vulnerabilities retain their potential of obtaining the Tapo account password of the victim user, even when exploited on other Tapo devices. This could allow the attacker to access the victim's account and control all associated devices. Additionally, the possibility to obtain the password of the victim's private network should not be underestimated as well, as network access can be the entry point for the attacker to execute different attacks on other devices within the network.

\bibliographystyle{apalike}
{\small
\bibliography{main}}
\end{document}